**Kolmogorov versus Iroshnikov-Kraichnan spectra: Consequences for ion heating in the solar wind**

C. S. Ng[1], A. Bhattacharjee[2], D. Munsi[2], P. A. Isenberg[2], and C. W. Smith[2]

[1] Geophysical Institute, University of Alaska Fairbanks, Fairbanks, Alaska, USA.

[2] Center for Magnetic Self-Organization, Center for Integrated Computation and Analysis of Reconnection and Turbulence, Institute for the Study of Earth, Ocean and Space, University of New Hampshire, Durham, New Hampshire, USA.

ABSTRACT

Whether the phenomenology governing MHD turbulence is Kolmogorov or Iroshnikov-Kraichnan (IK) remains an open question, theoretically as well as observationally. The ion heating profile observed in the solar wind provides a quantitative, if indirect, observational constraint on the relevant phenomenology. Recently, a solar wind heating model based on Kolmogorov spectral scaling has produced reasonably good agreement with observations, provided the effect of turbulence generation due to pickup ions is included in the model. Without including the pickup ion contributions, the Kolmogorov scaling predicts a proton temperature profile that decays too rapidly beyond a radial distance of 15 AU. In the present study, we alter the heating model by applying an energy cascade rate based on IK scaling, and show that the model yields higher proton temperatures, within the range of observations, with or without the inclusion of the effect due to pickup ions. Furthermore, the turbulence correlation length based on IK scaling seems to follow the trend of observations better.

# 1. Introduction

Since the proton temperature in the solar wind is observed to decrease with heliocentric distance slower than predicted by adiabatic expansion, it is believed that an *in situ* source is required to heat the solar wind [*Freeman*, 1988; *Richardson et al.*, 1995]. Good agreement with the observed temperature profile has been obtained using a quasi-steady solar wind turbulence evolution model [*Matthaeus et al.*, 1994; *Matthaeus et al.*, 1996; *Zank et al.*, 1996; *Matthaeus et al.*, 1999] that includes turbulence generation due to pickup ions in the outer heliosphere [*Williams et al.,* 1995; *Zank et al.*, 1996; *Smith et al.*, 2001; *Isenberg et al.*, 2003; *Smith et al.*, 2006]. Later developments of this model include extensions to the case with nonzero cross-helicity [*Matthaeus et al.*, 2004; *Breech et al.*, 2005, 2008].

The heating of the solar wind in the model is provided by the dissipation of turbulent energy, which cascades from large to small scales, and is eventually dissipated at the dissipation scale. Since in steady state, the heating rate is essentially the same as the energy cascade rate in the inertial range, the precise functional form of the cascade rate is an important ingredient of the model. Although different forms of the cascade rate were considered in the early development of the model [*Matthaeus et al.*, 1994; *Hossain et al.*, 1995], based on the Kolmogorov theory of hydrodynamic turbulence [*Kolmogorov*, 1941] as well as the Iroshnikov–Kraichnan (IK) theory of incompressible magnetohydrodynamic (MHD) turbulence [*Iroshnikov*, 1963; *Kraichnan*, 1966], later work involving detailed comparisons with observations was done assuming the Kolmogorov cascade rate. Whether anisotropic MHD turbulence should follow Kolmogorov or IK scaling remains an open question and a subject of significant ongoing research. We do not concern ourselves with this fundamental question here, but investigate the consequences of Kolmogorov or IK scaling as far as the problem of proton heating is concerned, assuming that the turbulence is isotropic.

We note that while there are many observations of the spectral index of solar wind turbulence that are considered more consistent with the Kolmogorov -5/3 value, instead of -3/2 of the IK theory [e. g, *Goldstein et al.,* 1995], there are still large enough uncertainties in these observed values that none of these theories can be ruled out definitively [for a recent review on solar wind turbulence, see *Bruno and Carbone*, 2005]. For examples, the observed values of the spectral index can change with time, location, and uncertainties regarding the precise extent of the inertial range. Recently, it has been reported that the spectral indices for velocity and magnetic fluctuations can be different, with velocity index closer to the IK value, and magnetic index closer to the Kolmogorov value [*Podesta et al*. 2006, 2007; *Tessein et al*., 2009]. However, as pointed out in [*Ng et al*., 2003], although the difference between 5/3 and 3/2 is small, and not unambiguously resolvable by observations, the energy cascade rate (and thus turbulent heating rate) predicted by these two theories can have significant differences, by an order of magnitude. Therefore, looking at the effects of turbulent heating might provide another way to distinguish between these two theories.

Since the solar wind heating model based on Kolmogorov scaling has already been shown to produce good agreement with the observed ion temperature profile, one might expect that using the IK energy cascade rate would not produce good agreement since it provides a much smaller heating rate compared with the Kolmogorov rate, if the level of turbulent fluctuations is held fixed. However, a recent study suggests that the solar wind heating rate at 1 AU is more consistent with the expectation from the IK cascade rate, and is about an order of magnitude smaller than expected from the Kolmogorov cascade rate [*Vasquez et al.*, 2007]. Since this study is carried out at one radial location, and the results are interesting as well as surprising, it is natural to ask what the results would be if the IK cascade rate is used instead of the Kolmogorov cascade rate in the solar wind heating model which attempts to make predictions of proton heating as a function of the radial distance. In this paper, we carry out this project. Specifically, we will repeat the calculations described in *Smith et al.* [2001], *Isenberg et al.* [2003] and *Smith et al.* [2006], except that all terms that depend on the Kolmogorov cascade rate are replaced by those based on the IK cascade rate. The new set of evolution equations are given in Section 2. The predictions of the new equations, and comparisons with calculations based on the Kolmogorov cascade rate, are given in Section 3. Discussion and conclusions will be presented in Section 4.

**2. Solar Wind Heating Model**

The solar wind heating model discussed in Section 1 is derived based on several strong and simplifying assumptions (see *Breech et al.* [2008] and other references therein). Among the principal assumptions are a steady and a spherically symmetric solar wind, an isotropic Kolmogorov scaling, a constant radial solar wind speed $V_{SW}$, and a constant Alfvén speed $V_A (\ll V_{SW})$. Under these conditions, the evolution of solar wind turbulence as a function of the heliocentric distance $r$ can be modeled by the following set of equations [*Smith et al.*, 2001, 2006; *Isenberg et al.*, 2003; *Isenberg*, 2005]:

$$\frac{dZ^2}{dr} = -\frac{A'}{r}Z^2 - \frac{\alpha Z^3}{\lambda V_{SW}} + \frac{Q}{V_{SW}} , \qquad (1)$$

$$\frac{d\lambda}{dr} = -\frac{C'}{r}\lambda + \frac{\beta Z}{V_{SW}} - \frac{\beta \lambda Q}{V_{SW} Z^2} , \qquad (2)$$

$$\frac{dT}{dr} = -\frac{4T}{3r} + \frac{m\alpha Z^3}{3k_B \lambda V_{SW}} . \qquad (3)$$

In this model, the turbulence is characterized by two quantities: the average fluctuation energy (in Elsässer units) $Z^2 = \langle \delta v^2 + \delta b^2/4\pi\rho \rangle$, where $\rho = nm$ is the solar wind density ($n$ and $m$ are proton density and mass respectively), and the correlation length of the fluctuations, $\lambda$. Note that by describing the turbulence energy by only one field, $Z^2$, we have assumed zero cross-helicity, i.e., $Z_+^2 = Z_-^2 = Z^2$, where $Z_\pm^2 = \langle \mathbf{Z}_\pm^2 \rangle$ with $\mathbf{Z}_\pm = \delta\mathbf{v} \pm \delta\mathbf{b}/(4\pi\rho)^{1/2}$. In this paper, we will concentrate on the case with zero cross helicity, although we have also obtained similar results by generalizing the set of equations to include nonzero cross-helicity. The constant parameters $A'$ (negative) and $C'$ are used to model the effect of stream compressions and shear. The function $Q$ (with the functional form given below) represents the fluctuation source due to interstellar pickup protons. The constant parameters $\alpha$ and $\beta$ are estimated from considerations of local turbulence theory [*Hossain et al.*, 1995; *Matthaeus et al.*, 1996]. The factor $Z^3/\lambda$ in the second term on the right hand side of Eq. (1) or (3) is due to the Kolmogorov cascade rate (see discussion below in this section). Here $T$ is the solar wind proton temperature, which evolves passively according to Eq. (3) but does not affect the evolution of $Z^2$ and $\lambda$. Note that the first term on the right hand side of Eq. (3) describes the adiabatic cooling due to the expansion of the solar wind, while the second term represents the heating due to dissipation of the turbulent energy.

As pointed out above, this set of equations is based on the assumption of a Kolmogorov cascade. In the Kolmogorov theory, the energy $\delta v^2$ of the scale $\lambda$ is estimated to cascade to the next scale in an eddy turnover time $\tau \sim \lambda/\delta v$. Therefore, the energy cascade rate is $\varepsilon \sim \delta v^2/\tau \sim \delta v^3/\lambda$. On the other hand, in the IK theory of MHD turbulence, the energy cascade is inhibited by the fact that an Alfvén wave packet moving in one direction does not cascade energy to smaller scales except when it collides with another Alfvén wave packet moving in the opposite direction. However, this collision time $\tau_A \sim \lambda/V_A$ is much smaller than the eddy turnover time $\tau \sim \lambda/\delta v$ so that it takes many random collisions to cascade the same amount of energy. In fact, the energy cascade time can be estimated to be $\tau_E \sim \tau^2/\tau_A \sim \lambda V_A/\delta v^2$, and thus $\varepsilon \sim \delta v^2/\tau_E \sim \delta v^4/\lambda V_A$. In order to examine the effect of using IK cascade, we rewrite Eqs. (1)-(3), as follows:

$$\frac{dZ^2}{dr} = -\frac{A'}{r}Z^2 - \frac{\alpha Z^4}{\lambda V_{SW}V_A} + \frac{Q}{V_{SW}} \quad , \tag{4}$$

$$\frac{d\lambda}{dr} = -\frac{C'}{r}\lambda + \frac{\beta}{V_{SW}}\left(\frac{Z^4}{V_A}\right)^{1/3} - \frac{\beta\lambda Q}{V_{SW}Z^2} \quad , \tag{5}$$

$$\frac{dT}{dr} = -\frac{4T}{3r} + \frac{m\alpha Z^4}{3k_B\lambda V_{SW}V_A} \quad . \tag{6}$$

Note that the factor $Z^4/\lambda V_A$ in the second term of the right hand side of Eq. (4) or (6) is due to the IK energy cascade rate. The form of the second term on the right hand side of Eq. (5) is due to *Matthaeus et al.* [1994] and *Hossain et al.* [1995]. Although the pickup ions terms involving $Q$ appear formally unchanged, they too depend on the assumed energy cascade rate, as described below.

The function $Q$ in this model is calculated using the expression

$$Q = \zeta \frac{V_{SW}^2}{n} \frac{dN}{dt}, \qquad (7)$$

where $\zeta$, to be determined later, is the fraction of newly ionized pickup proton energy that generates waves. Here $V_{SW}^2/n$ is the initial kinetic energy per pickup proton in the same units as $Z^2$ in the plasma frame, and $dN/dt$ is the rate at which pickup protons are created, which can be modeled by the equation $dN/dt = N_0 \nu_0 (r_E/r)^2 \exp(-L/r)$, where $L$ is the scale of the ionization cavity, $N_0$ is the neutral hydrogen density at the termination shock, and $\nu_0$ is the ionization rate at $r = r_E = 1$ AU.

Following *Isenberg et al.* [2003] and *Isenberg* [2005], the factor $\zeta$ is calculated from the equation

$$\zeta(\Delta) = 1 - \frac{\Delta + V_{SW}^{-4} \int_\Delta^1 v^4(\mu) S(\mu) d\mu}{\Delta + V_{SW}^{-2} \int_\Delta^1 v^2(\mu) S(\mu) d\mu}, \qquad (8)$$

where $\Delta = Z/3^{1/2} V_A$, $v(\mu)$ is the solution obtained by integrating the equation,

$$\frac{dv}{d\mu} = \left[ \sum_j \frac{V_j I_j(k_r)}{|\mu v - W_j|} \left(1 - \frac{\mu V_j}{v}\right) \right] \left[ \sum_j \frac{I_j(k_r)}{|\mu v - W_j|} \left(1 - \frac{\mu V_j}{v}\right)^2 \right]^{-1}, \qquad (9)$$

subject to the initial condition of $v(\mu = \Delta) = V_{SW}$, $S(\mu)$ is a scale factor calculated by taking the difference between $v(\mu)$ and another solution of Eq. (9) using the initial condition $v(\mu = \Delta) = 1.001 V_{SW}$, normalized to $S(\mu = \Delta) = 1$. Here $V_j$ and $W_j$ in Eq. (9) are the phase and group velocity of the *j*th wave mode resonating with the cyclotron resonant wave number

$$k_r = \frac{\Omega}{\mu v - V_j}, \qquad (10)$$

where $\Omega$ is the proton cyclotron frequency. Using the cold plasma dispersion relation

$$\omega(k) = \pm k V_A \sqrt{1 + \frac{\omega}{\Omega}} \ , \tag{11}$$

$V_j$ can be obtained by solving the third-order equation

$$V_j^3 - \mu v V_j^2 + \mu v V_A^2 = 0 \ , \tag{12}$$

and $W_j$ is given by

$$W_j = -\frac{2\mu v V_A^2}{2V_j^2 - 2\mu v V_j + V_A^2} \ . \tag{13}$$

Note that there is only one resonant wave mode if $|\mu v| < 1.5\sqrt{3} V_A$, and three modes otherwise. The function $I(k)$ in Eq. (9) is determined by the one-dimensional energy spectrum of the turbulence. When the energy spectrum is Kolmogorov, $I(k)$ is given by

$$I(k) = A(r)|k|^{-5/3} \ . \tag{14}$$

Note that the function $A(r)$ does not enter the final result since it is cancelled in Eq. (9) at each position $r$. For the present study using the IK scaling using Eqs. (4)-(6), we need to use the IK spectrum instead, i.e.,

$$I(k) = A(r)|k|^{-3/2} \ . \tag{15}$$

Note that the above formulation to calculate $\zeta$ follows *Isenberg* [2005], which is a corrected version of the analysis in *Isenberg et al.* [2003]. The correction was shown in *Isenberg* [2005] to change the resulting solar wind temperature only slightly, and both agree well with observations.

The coefficients $A'$ and $C'$ in these two sets of equations can in principle be different, depending on the spectral index, and this variation may change the model predictions significantly. However, since we estimate them by dimensional arguments [e.g., see *Breech et al.*, 2008], which do not depend on the spectral index explicitly, we will choose values for $A'$ and $C'$ that are the same as those used in previous studies [*Smith et al.*, 2001, 2006; *Isenberg et al.*, 2003; *Isenberg*, 2005], in order to have a meaningful comparison with earlier results.

### 3. Numerical Results

We now present numerical results obtained by solving Eqs. (4)-(6), with the $Q$ term calculated using the IK scaling (15). In order to compare with previous results obtained by *Smith et al.* [2001] and *Isenberg et al.* [2003], based on Eqs. (1)-(3) with Kolmogorov scaling, we will use the same parameters as in these two earlier papers, summarized in Table 1. Also, Fig. 1 and Fig. 2 are plotted in formats very close to corresponding figures in these papers (i.e., Fig. 5 in [*Isenberg et al.*, 2003] and Fig. 7 in [*Smith et al.,* 2001]) for ease of comparison.

In the first case, based on parameters used in *Isenberg et al.* [2003], results on the proton temperature are plotted in Fig. 1. The fluctuating curve is the running average of the solar wind temperature measured over 51 days by *Voyager 2* versus the heliocentric distance $r$. The purple dashed curve represents the prediction of temperature with only adiabatic cooling, i.e., only keeping the first term of the right hand side of Eq. (3). As is well known, this prediction is much lower than the observed temperature. This indicates the need for including a heating source in the model, e.g., the turbulence cascade, represented by the second term on the right hand side of Eq. (3) or (6).

The solid black curve in Fig. 1 is the model temperature predicted by Eqs. (1)-(3), based on the Kolmogorov scaling (14), as calculated by *Isenberg et al.* [2003], including the effect of pickup protons. We see that the prediction of the model agrees well with observations. Such good agreement between observations and model predictions based on the Kolmogorov theory can perhaps be interpreted by some as a confirmation of the correctness of the Kolmogorov scaling in solar wind turbulence, but our results indicate that this is not the only possible conclusion.

The green curve is obtained by again using Eqs. (1)-(3), but with $Q$ set to zero so that we may see the effect of pickup ions. This curve is basically the same as the black curve up to around 10 AU, since the effect of pickup ions is only significant in the outer heliosphere. Beyond this distance, we see that the prediction without the $Q$ term would be significantly lower than observations suggest, and does not show the trend of increasing temperature beyond around 20 AU. We thus see that the effect of the pickup ions in this model is indeed very important in obtaining good predictions of solar wind temperature in the outer heliosphere.

The blue curve is calculated from our model based on the IK cascade, i.e., Eqs. (4)-(6) with $Q = 0$. We see that the predictions of the proton temperature in this case are significantly larger than those given by the green curve beyond around 5 AU. In fact, the predictions given by the blue curve appear to be consistent with observations despite the exclusion of pickup ions, and fall below the observed temperature only beyond about 40 AU.

Since the black curve with $Q$ is significantly higher than the green curve for $Q = 0$, one might expect that adding the effect of pickup ions to the IK cascade rate will overestimate the proton temperature when compared with observations. However, somewhat surprisingly, the red curve, which is obtained from Eqs. (4)-(6) with $Q$ determined using the IK scaling (15), is seen to lie only slightly above the blue curve.

This shows that the effect of turbulence generation due to pickup ions is weaker when the IK spectrum (15) is used instead of the Kolmogorov spectrum (14). We can understand this by considering the physics of the $Q$ term, which is essentially a measure of whether pickup ions give energy to a spectrum of waves (positive $Q$), or gain energy from it (negative $Q$). A pickup ion gives energy when it interacts with a backward moving wave at smaller wave number $k$, but gains energy when it interacts with a forward moving wave at larger $k$, due to the Doppler effect as indicated in Eq. (10) (see also [*Isenberg et al.*, 2003; *Isenberg*, 2005]). The strength of such interactions is proportional to the intensity of the waves. For a turbulent spectrum of waves that decrease in intensity at larger values of $k$, pickup ions give energy to waves and result in a positive $Q$ term. So, when the IK spectrum, which is flatter in $k$ space, is used, there is a stronger cancellation between the two effects, resulting in a smaller value of $Q$.

We have mentioned that the IK cascade rate is smaller than the Kolmogorov cascade rate, for the same level of turbulence. In view of this, the above results, which show that the IK scaling actually produces a higher temperature, might seem counter-intuitive. To understand this physically, we also need to look at the comparisons of $Z^2$ and $\lambda$. We will now do so by repeating the calculation in *Smith et al.* [2001] using our model based on the IK cascade, since the plots of $Z^2$ and $\lambda$ in [*Isenberg et al.*, 2003] only have model outputs without observation data.

For the second case based on parameters used in *Smith et al.* [2001], results are plotted in Fig. 2, showing (a) $Z^2$ normalized to the value at 1 AU, i.e., $Z^2(r)/Z^2(1\ \mathrm{AU})$, versus the heliocentric distance $r$, (b) the correlation length $\lambda$, and (c) the temperature $T$ normalized to the value at 1 AU, i.e., $T(r)/T(1\text{ AU})$. The discrete data points are from *Voyager 2* observations, the same as those used in the first case.

The solid green curves are the model predictions obtained by using Eqs. (1)-(3) with $Q = 0$. The black curves are from the same set of equations including the effect of pickup protons. Like the results in *Isenberg et al.* [2003], the temperature predictions agree reasonably well with observations when the $Q$ term is taken into account. Without this contribution, the temperature falls significantly below observations in the outer heliosphere. At the same time, the predictions of $Z^2$, with or without $Q$, are consistent with observations, with the predictions for nonzero $Q$ slightly higher than those with $Q = 0$. However, the predictions of the model for the correlation length $\lambda$ are not very good. When $Q$ is set to zero, $\lambda$ does appear to follow qualitatively the overall trend of the observed data, although it is somewhat lower in values. When the effect of $Q$ is included, the predictions of $\lambda$ deviate from observations even more. In fact, the predictions turn around and decrease with *r* at larger distance, opposite to the qualitative trend seen in the observations. (See *Smith et al*. [2001] for further discussions of this discrepancy).

The blue curves are calculated from our model based on IK cascade, i.e., Eqs. (4)-(6) with $Q = 0$. As in the first case, we see that the temperature predictions already are well within the range of observations for all distances. The predictions of $Z^2$ are also consistent with observations, in contrast to those given by the green or the black curve. The predictions of $\lambda$ also show a similar trend.

The red curves are obtained from the same set of equations, with $Q$ calculated using the IK scaling (15). We see again that the effect of pickup ions is not very significant when the IK scaling is used. Therefore, predictions of all three quantities (red curves) differ from the respective blue curves only slightly. The predictions for the temperature do show a slight trend of increasing with $r$ when the effect of $Q$ is included. The predictions of $Z^2$ deviate upward from those of the $Q=0$ case noticeably only for very large $r$. On the other hand, the predictions of $\lambda$ move downward from those of the $Q=0$ case, moving closer to observations. This differs significantly from the predictions using the Kolmogorov scaling with finite $Q$, since now the model does predict the correct trend of increasing $\lambda$ with $r$.

Note that the model curves in the temperature plot, i.e., Fig. 2(c), are slightly different from those in Fig. 1, due to differences in parameters used, as indicated in Table 1, not because of any difference in model methods or observational data.

One unexpected result in the present study is that the heating provided by the IK cascade is actually at the same level or even greater than that obtained by using the Kolmogorov cascade. Using quantities defined in this paper, the Kolmogorov cascade rate is given by $\varepsilon_K \sim Z^3/\lambda$, which is formally greater than the IK cascade rate $\varepsilon_{IK} \sim Z^4/\lambda V_A$ for the same level of $Z$ and $\lambda$ (since $Z \ll V_A$ usually). From the results in Fig. 2, we see that the reason why $\varepsilon_{IK}$ is of the same order, or greater than $\varepsilon_K$ (and thus the solar wind temperature predictions are roughly the same) is that the turbulence level $Z^2$ predicted by the model using IK scaling is greater than that predicted by using Kolmogorov scaling, sometimes by more than an order of magnitude. (This is despite the fact that the predicted $\lambda$ using IK scaling is also greater than that by using Kolmogorov scaling, but only by about a factor of three for the $Q=0$ case.) The reason for this behavior can be traced back to the turbulence generation term, i.e., the first term on the right hand side of Eq. (1) or (4). At $r$ around 1 AU, $Z^2$ in the Kolmogorov and IK runs are about the same. However, due to the fact that the $\varepsilon_K$ is greater than $\varepsilon_{IK}$ at first, $Z^2$ decreases in the Kolmogorov case faster than that in the IK case. This effect is amplified by the fact that a smaller $Z^2$ generates less turbulence (note that $A' < 0$). This is also why the solar wind temperature in the Kolmogorov case is slightly larger than that in the IK case around 1 AU, up to about 3 AU. However, as this effect continues, $\varepsilon_{IK}$ begins to catch up with $\varepsilon_K$ when $Z^2$ in the Kolmogorov case is much smaller than that in the IK case, and thus the temperature predictions by the two runs become roughly the same. At larger $r$, $\varepsilon_{IK}$ is actually larger than $\varepsilon_K$ when $Q = 0$. $\varepsilon_K$ gets back to about the same level as $\varepsilon_{IK}$ only after we include the effect of pickup ions, since this effect is stronger when the Kolmogorov spectrum is used. Thus, after all these effects are taken into account, the predictions of the solar wind temperature are roughly the same in the two cases, despite the fact that the turbulence cascade rates of the two theories are very different.

Finally, we also consider the case presented in *Smith et al.* [2006], who use a more direct method of comparing predictions from the mode equations (1)-(3) with observations. In the cases discussed above, the boundary conditions on $Z^2$, $\lambda$, and *T* at *r* = 1 AU are held fixed in obtaining predictions for all *r*. However, the predictions of the model are being compared with observations obtained at different positions and necessarily at different times, since the data are obtained from the same steadily moving spacecraft as it moves out towards the outer heliosphere. Therefore, an implicit assumption of the earlier studies is that the solar wind conditions are quasi-steady. However, this assumption is not generally true. To get around this difficulty, *Smith et al.* [2006] used the observed solar wind speed (which is assumed to be constant) at different positions *r* to determine when that fluid element actually passed through 1 AU. Then the solar wind conditions at that time at 1 AU are determined using Omnitape data, and used as boundary conditions for Eqs. (1)-(3). More detailed description of this method can be found in *Smith et al.* [2006]. Here we follow their method, and repeat our study.

In Fig. 3(a), the red curve is the proton temperature $T_p$ in K as a function of heliocentric distance in AU, calculated from Eqs. (1)-(3) using the method and parameters used in *Smith et al.* [2006]. The discrete data points are from *Voyager 2* observations. We see that the predictions from the model are consistent with observations until about 43 AU. From there to about 55 AU, the predictions are substantially lower than observations. This is identified by *Smith et al.* [2006] as a latitude effect, since *Voyager 2* was at high latitude. The predictions beyond 55 AU are also found to be somewhat higher than observations (about a factor of two on the average). In 3(b), the $T_p$ curve is now calculated from Eqs. (4)-(6) using the same parameters. We see that the agreement with observations up to about 43 AU is about the same as in the case (a). At the same time, the predictions beyond 55 AU are now lower, consistent with observations. However, the discrepancy with data from 43 to 55 AU is worse. However, since the main discrepancy in this region is due to the high-latitude effect, it is hard to separate out the effects due to turbulence spectral laws. For the IK case, we also do two more test runs. In 3(c), we repeat the run as in (b), but artificially set $Q$ to zero. We then see that predictions beyond 55 AU become lower and less consistent with observations, although not by much. Overall, this seems to suggest that the pickup ion term does possibly provide important corrections, although these corrections are not as crucial for IK scaling as they are for Kolmogorov scaling. To reinforce this point, we run the case in 3(d), where we repeat the run (b) but with the $Q$ term calculated using a spectrum with a spectral index of 5/3 (Kolmogorov) rather than 3/2 (IK). For this case, we see that the predictions for the proton temperature are slightly higher than that in (b) for *r* beyond 55 AU, but not as high as the case in (a). This suggests that the most important effect of change from Kolmogorov to IK scaling is due to other terms in Eqs. (4)-(6), rather than the *Q* term.

**4. Summary and Conclusions**

In this paper, we have investigated the effect of turbulence scaling laws on the heating of solar wind by substituting the IK cascade rate into a solar wind turbulence evolution model, replacing the Kolmogorov cascade rate, and comparing with observational results from *Voyager 2* on the solar wind temperature, turbulence energy level, and correlation length. The surprising result of this study is that the solar wind temperature predicted by using the IK cascade is comparable with that using the Kolmogorov cascade. This is true whether the effect of pickup ions (the *Q* term) is included or not (including the pickup ions term does seem to give slightly better results), since we show that the effect of pickup ions is weaker when the IK spectrum is used than when the Kolmogorov spectrum is used. The reason for this is principally due to the fact that the turbulence energy level ($Z^2$) in the IK case decays more slowly than that in the Kolmogorov case as we move out radially in the heliosphere. The predictions on the correlation length ($\lambda$) in the IK case are also consistent with observations, with or without the pickup ions. This is in contrast with the Kolmogorov case, which has the correct trend only when the effect of pickup ions is excluded, but shows a qualitative discrepancy with the data when the effect of pickup ions is included.

Since this solar wind turbulence evolution model is based on drastic assumptions and the observations have significant uncertainties, the fact that the model using either cascade law has predictions that are consistent with observations does not necessarily confirm the correctness of either scaling law. However, from the present study, we do see that the IK theory produces at least as good a comparison with observations as the Kolmogorov theory. More theoretical as well as observational investigations are necessary to distinguish between the consequences for each phenomenological theory to solar wind turbulence.

There are at least two ways to improve the existing model: include the effects of cross-helicity in the presence of the IK cascade, and the effects of anisotropy. We plan to investigate these effects in the future.

**Acknowledgements**

This research is supported in part by National Science Foundation Grant No. AST-0434322, National Aeronautics and Space Administration Grant No. NNX08BA71G, and the Department of Energy under the auspices of the Center for Magnetic Self-Organization (CMSO) and the Center for Integrated Computation and Analysis of Reconnection and Turbulence (CICART).

FIGURE CAPTIONS

Figure 1. The fluctuating curve is the 51-days running average solar wind temperature measured at *Voyager 2* versus the heliocentric distance *r*. The solid black curve is the model temperature calculated by Eqs. (1)-(3), using parameters of *Isenberg et al.* [2003], including the effect of pickup protons. The green curve is using the same model as in *Isenberg et al*. [2003], i.e., Eqs. (1)-(3), with $Q = 0$. The blue curve is calculated from Eqs. (4)-(6) with $Q = 0$, while the red curve is from the same set of equations with $Q$ calculated using the IK scaling of Eq. (15). The purple dashed curve is simply the prediction of temperature with only adiabatic cooling, i.e., only keeping the first term of the right hand side of Eq. (3).

Figure 2. (a) $Z^2$ normalized to the value at 1 AU versus the heliocentric distance *r*; (b) the correlation length $\lambda$; (c) temperature *T* normalized to the value at 1 AU. The discrete data points are from *Voyager 2* observations. The green curves are the model predictions calculated by Eqs. (1)-(3) with $Q = 0$. The black curves are from the same set of equations including the effect of pickup protons. The blue curve is calculated from Eqs. (4)-(6) with $Q = 0$, while the red curve is from the same set of equations with $Q$ calculated using the IK scaling of Eq. (15).

Figure 3. (a) The red curve is solar wind proton temperature $T_p$ in K as a function of heliocentric distance in AU calculated from Eqs. (1)-(3) using the method and parameters used in *Smith et al*. [2006]. The discrete data points are from *Voyager 2* observations. (b) The $T_p$ curve is now calculated from Eqs. (4)-(6) using the same parameters. (c) Same as

(b) but with $Q = 0$. (d) Same as (b) but with the $Q$ term calculated using a spectrum with a spectral index of 5/3 (Kolmogorov) rather than 3/2 (IK).

**References**

Breech, B., Matthaeus, W. H., Minnie, J., Bieber, J. W., Oughton, S., Smith, C. W., & Isenberg, P. A. (2008), *J. Geophys. Res.*, *113*, A08105, doi:10.1029/2007JA012711.

Breech, B., W. H. Matthaeus, J. Minnie, S. Oughton, S. Parhi, J. W. Bieber, and B. Bavassano (2005), Radial evolution of cross helicity in high latitude solar wind, *Geophys. Res. Lett.*, *32*, L06103, doi:10.1029/2004GL022321.

Bruno, R., and V. Carbone (2005), The solar wind as a turbulence laboratory, *Living Rev. Solar Phys., 2*, 4-186.

Goldstein, B. E., Smith, E. J., Balogh, A., Horbury, T. S., Goldstein, M. L., Roberts, D. A., (1995), Properties of magnetohydrodynamic turbulence in the solar wind as observed by Ulysses at high heliographic latitudes, *Geophys. Res. Lett., 22*, 3393-3396.

Freeman, J. W. (1988), Estimates of solar wind heating inside 0.3 AU, *Geophys. Res. Lett., 15*, 88-91.

Hossain, M., P. C. Gray, D. H. Pontius Jr., W. H. Matthaeus, and S. Oughton (1995), Phenomenology for the decay of energy-containing eddies in homogeneous turbulence, *Phys. Fluids*, *7*, 2886–2904.

Iroshnikov, P. S. (1963), *Astron. J. SSSR, 40*, 742-750; Turbulence of a conducting fluid in a strong magnetic field, (*Soviet Astron. 7*, 566, 1963).

Isenberg, P. A. (2005), Turbulence-driven solar wind heating and energization of pickup protons in the outer heliosphere, *Astrophys. J., 623*, 502–510.

Isenberg, P. A., C. W. Smith, and W. H. Matthaeus (2003), Turbulent heating of the distant solar wind by interstellar pickup protons, *Astrophys. J., 592*, 564–573.

Kolmogorov, A. N. (1941), The local structure of turbulence in incompressible viscous fluid for very large Reynolds numbers, *Dokl. Akad. Nauk. SSSR, 30*, 301-305.

Kraichnan, R. H. (1965), Inertial-range spectrum of hydromagnetic turbulence, *Phys. Fluids, 8*, 1385-1387.

Matthaeus, W. H., J. Minnie, B. Breech, S. Parhi, J. W. Bieber, and S. Oughton (2004), Transport of cross helicity and the radial evolution of Alfvénicity in the solar wind, *Geophys. Res. Lett., 31*, L12803, doi:10.1029/2004GL019645.

Matthaeus, W. H., S. Oughton, D. Pontius, and Y. Zhou (1994), Evolution of energy containing turbulent eddies in the solar wind, *J. Geophys. Res., 99*, 19,267–19,287.

Matthaeus, W. H., G. P. Zank, and S. Oughton (1996), Phenomenology of hydromagnetic turbulence in a uniformly expanding medium, *J. Plasma Phys., 56*, 659–675.

Matthaeus, W. H., G. P. Zank, C. W. Smith, and S. Oughton (1999), Turbulence, spatial transport, and heating of the solar wind, *Phys. Rev. Lett., 82*, 3444–3447.

Ng, C. S., A. Bhattacharjee, K. Germaschewski, and S. Galtier (2003), Anisotropic fluid turbulence in the interstellar medium and the solar wind, *Phys. Plasmas, 10*, 1954-1962.

Podesta, J. J., D. A. Roberts, and M. L. Goldstein (2006), Power spectrum of small-scale turbulent velocity fluctuations in the solar wind, *J. Geophys. Res., 111*, A10109, doi: 10.1029/ 2006JA011834.

Podesta, J. J., D. A. Roberts, and M. L. Goldstein (2007), Spectral Exponents of Kinetic and Magnetic Energy Spectra in Solar Wind Turbulence, *Astrophys. J., 664*, 543-548.

Richardson, J. D., K. I. Paularena, A. J. Lazarus, and J. W. Belcher (1995), Radial evolution of the solar wind from IMP 8 to Voyager 2, *Geophys. Res. Lett., 22*, 325–328.

Smith, C. W., P. A. Isenberg, W. H. Matthaeus, and J. D. Richardson (2006), Turbulent heating of the solar wind by newborn interstellar pickup protons, *Astrophys. J., 638*, 508–517.

Smith, C. W., W. H. Matthaeus, G. P. Zank, N. F. Ness, S. Oughton, and J. D. Richardson (2001), Heating of the low-latitude solar wind by dissipation of turbulent magnetic fluctuations, *J. Geophys. Res., 106*, 8253–8272.

Tessein, J. A., C. W. Smith, B. T. MacBride, W. H. Matthaeus, M. A. Forman, and J. E. Borovsky (2009), Spectral indices for multi-dimensional interplanetary turbulence at 1 AU, *Astrophys. J., 692*, 684-693.

Vasquez, B. J., C. W. Smith, K. Hamilton, B. T. MacBride, and R. J. Lemon (2007), Evaluation of the turbulent energy cascade rates from the upper inertial range in the solar wind at 1 AU, *J. Geophys. Res., 112*, A07101.

Williams, L. L., G. P. Zank, and W. H. Matthaeus (1995), Dissipation of pickup-induced waves: A solar wind temperature increase in the outer heliosphere? *J. Geophys. Res., 100*, 17,059–17,067.

Zank, G. P., W. H. Matthaeus, and C. W. Smith (1996), Evolution of turbulent magnetic fluctuation power with heliospheric distance, *J. Geophys. Res., 101*, 17,093–17,107.

TABLE 1

Parameters and boundary conditions used in Fig. 5 of *Isenberg et al*. [2003], and Fig. 7 of *Smith et al*. [2001].

| Parameter | Isenberg et al. [2003] | Smith et al. [2001] |
| --- | --- | --- |
| $V_{SW}$ (km/s) | 440 | 400 |
| $V_A$ (km/s) | 33 | 50 |
| $N_0$ (cm$^{-3}$) | 0.1 | 0.1 |
| $\nu_0$ (s$^{-1}$) | $7.5 \times 10^{-7}$ | $10^{-6}$ |
| $u_0$ (km/s) | 20 | 20 |
| $L$ (AU) | 5.6 | 8 |
| $A'$ | -1.1 | -1.1 |
| $C'$ | 1.8 | 1.8 |
| $\alpha$ | 1 | 1 |
| $\beta$ | 1 | 1 |
| $Z^2$(1 AU) [(km/s)$^2$] | 700 | 350 |
| $\lambda$(1 AU) (AU) | 0.03 | 0.03 |
| $T$(1 AU) (K) | $7 \times 10^4$ | $7 \times 10^4$ |

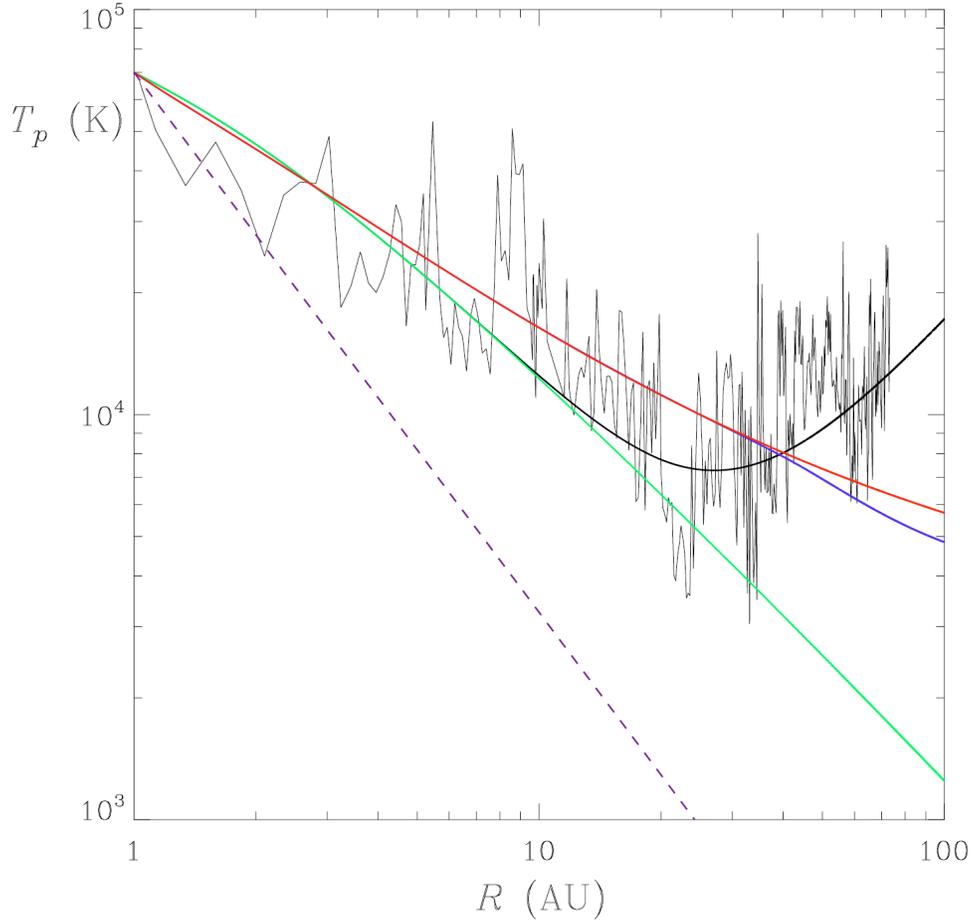

Figure 1. The fluctuating curve is the 51 days running average solar wind temperature measured at *Voyager 2* versus the heliocentric distance *r*. The solid black curve is the model temperature calculated by Eqs. (1)-(3), using parameters of *Isenberg et al.* [2003], including the effect of pickup protons. The green curve is using the same model as in *Isenberg et al.* [2003], i.e., Eqs. (1)-(3), with $Q$ being set to zero, to see the effect of the pickup ions term. The blue curve is calculated from Eqs. (4)-(6) with $Q = 0$, while the red curve is from the same set of equations with $Q$ calculated using the IK scaling of Eq. (15). The purple dashed curve is simply the prediction of temperature with only adiabatic cooling, i.e., only keeping the first term of the right hand side of Eq. (3).

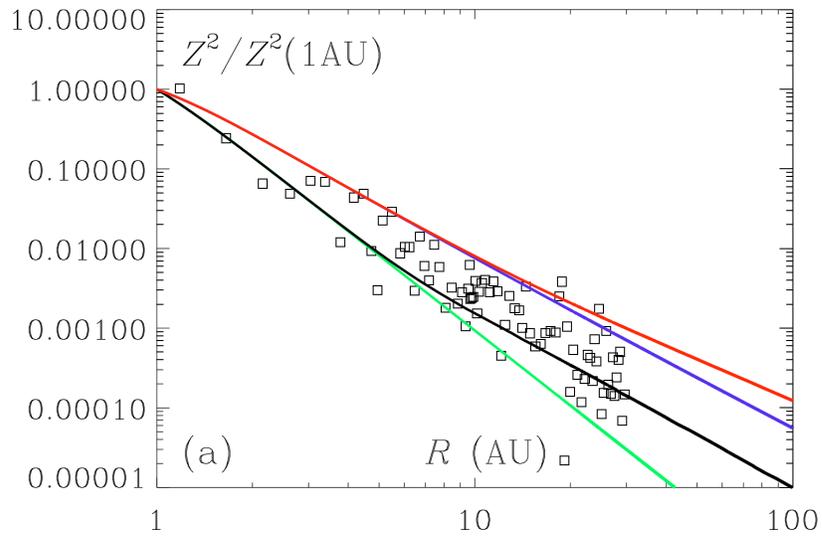

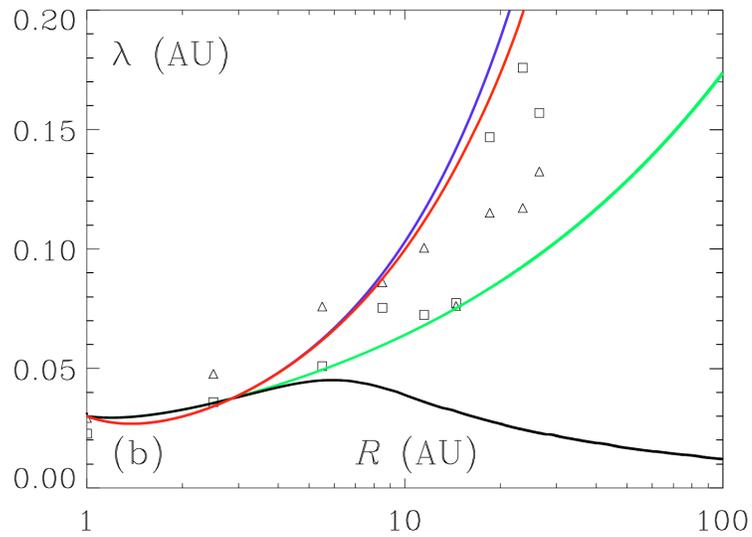

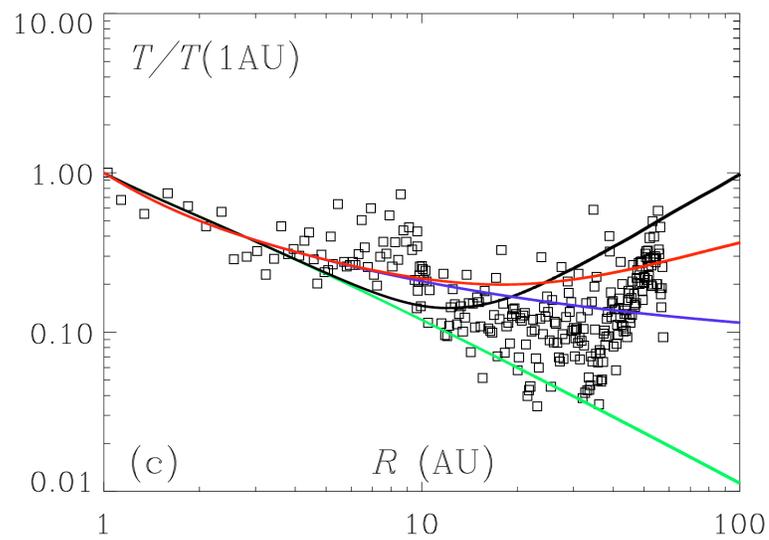

Figure 2. (a) $Z^2$ normalized to the value at 1 AU versus the heliocentric distance $r$; (b) the correlation length $\lambda$; (c) temperature $T$ normalized to the value at 1 AU. The discrete data points are from *Voyager 2* observations. The green curves are the model predictions calculated by Eqs. (1)-(3) with $Q=0$. The black curves are from the same set of equations including the effect of pickup protons. The blue curve is calculated from Eqs. (4)-(6) with $Q=0$, while the red curve is from the same set of equations with $Q$ calculated using the IK scaling (15).

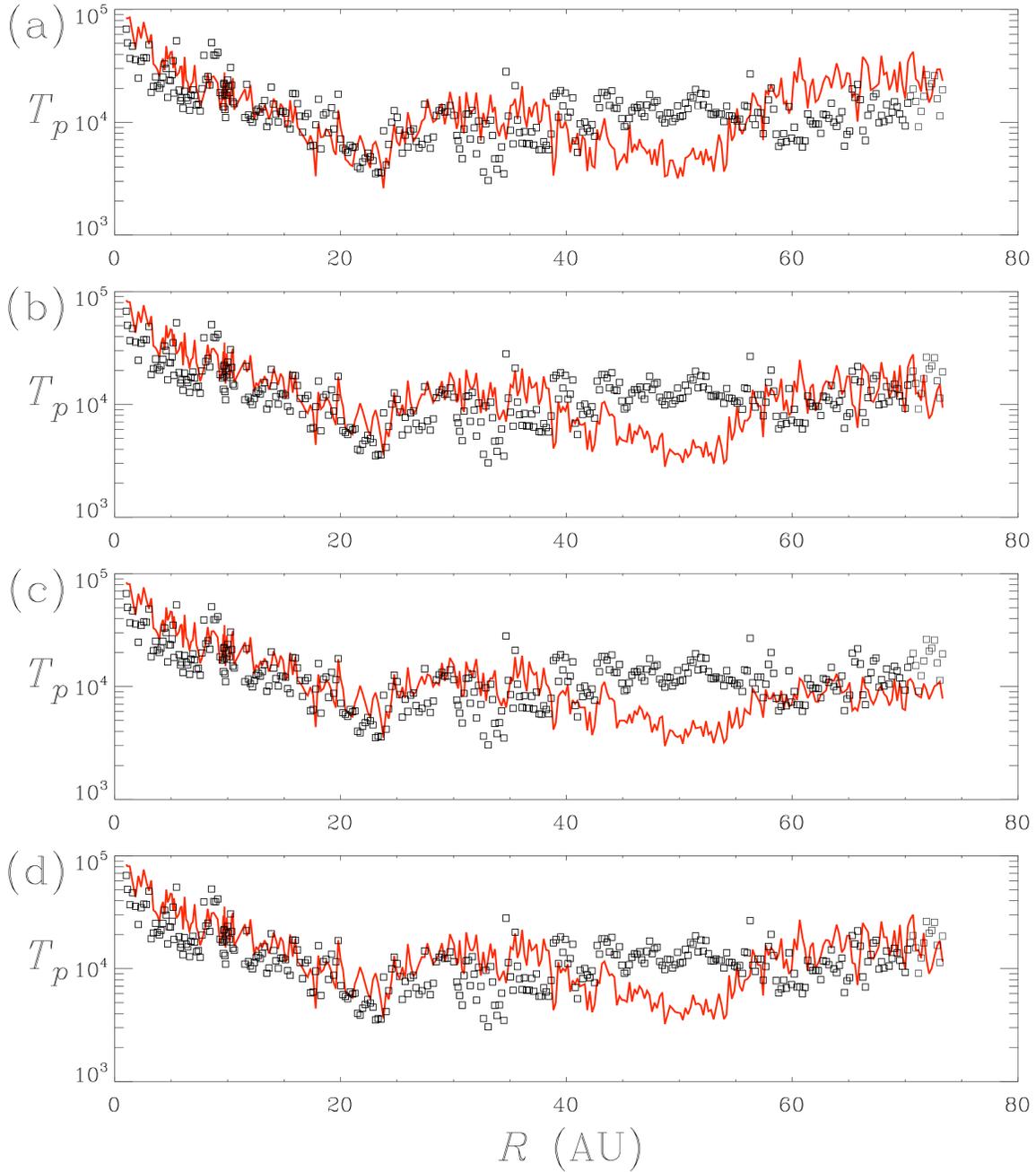

Figure 3. (a) The red curve is solar wind proton temperature $T_p$ in K as a function of heliocentric distance in AU calculated from Eqs. (1)-(3) using the method and parameters used in *Smith et al*. [2006]. The discrete data points are from *Voyager 2* observations. (b) The $T_p$ curve is now calculated from Eqs. (4)-(6) using the same parameters. (c) Same as (b) but with $Q$ set to zero. (d) Same as (b) but with the $Q$ term calculated using a spectrum with a spectral index of 5/3 (Kolmogorov) rather than 3/2 (IK).